\documentclass{article}
\usepackage{amssymb}
\usepackage{amsmath}
\usepackage{latexsym}
\usepackage{array}
\usepackage{graphicx}
\usepackage{bm}
\usepackage{exscale}

\usepackage{theorem}
\theoremstyle{break}
\newtheorem{Proposition}{Proposition}

\newtheorem{Definition}{Definition}
\def\qed{\hfill\hbox{$\Box$}\vspace{10pt}\break}

\def\C{{\mathbb C}}
\def\Z{{\mathbb Z}}
\def\Q{{\mathbb Q}}
\def\F{{\mathbb F}}
\def\R{{\mathbb R}}
\def\P{{\mathbb P}}

\def\II{${}_{\mbox{\scriptsize{II} }}$}

\begin{document}
\title{Discrete Painlev\'{e} II equation over finite fields}
\author{M Kanki$^1$, J Mada$^2$, K M Tamizhmani$^3$ and T Tokihiro$^1$\\
\small $^1$ Graduate School of Mathematical Sciences,\\
\small University of Tokyo, 3-8-1 Komaba, Tokyo 153-8914, Japan\\
\small $^2$ College of Industrial Technology,\\
\small Nihon University, 2-11-1 Shin-ei, Narashino, Chiba 275-8576, Japan\\
\small $^3$ Department of Mathematics,\\
\small Pondicherry University, Kalapet, 605014 Puducherry, India}
\date{}
\maketitle

\begin{abstract}
We investigate the discrete Painlev\'{e} II equation over finite fields.
We treat it over local fields and observe that it has a property that is similar to the good reduction over finite fields.
We can use this property, which seems to be an arithmetic analogue of singularity confinement, to avoid the indeterminacy of the equations over finite fields and to obtain special solutions from those defined originally over fields of characteristic zero.
{\tt PACS2010: 45.05.+x, 02.30.Ik}
\end{abstract}

\section{Introduction}
\label{sec1}
Dynamical systems are widely used as models for phenomena in natural or social science.
A dynamical system consists of a set and a mapping defined on it.
In Hamiltonian mechanics, this set is a vector space over a field of real numbers $\R^n$ and the mapping is defined by differential equations \cite{Arnold}.  
This is a typical example of a continuous dynamical system. 
On the other hand, a cellular automaton is an example of a discrete dynamical system, in which the set consists of a countable number of elements and the mapping is given by recurrence formulae or difference equations \cite{Wolfram}.
An interesting problem is whether the nature of the system significantly changes or not when the underlying set is changed to a different one.
For example, the tent map over $\R$ is a simple discrete dynamical system which exhibits chaotic behaviour for a generic initial state,
 however it only shows periodic behaviour if it is defined over $\Q$ \cite{Collet}.
This problem is of particular importance in so called arithmetic dynamics, which concerns the dynamics over arithmetic sets such as $\mathbb{Z}$ or $\mathbb{Q}$ or a number field that is of number theoretic interest \cite{Silverman}.
In arithmetic dynamics, the change of dynamical properties of polynomial or rational mappings give significant information when reducing them modulo prime numbers. 
The mapping is said to have good reduction if, roughly speaking, the reduction commutes with the mapping itself \cite{Silverman}.
Linear fractional transformations in $\mbox{PGL}_2$ are typical examples of mappings with good reduction.
Recently bi-rational mappings over finite fields have been investigated in terms of integrability
\cite{Roberts}.
The QRT mappings \cite{QRT} over finite fields have been studied in detail by choosing the parameter values so that indeterminate points are avoided \cite{Roberts2}.
They have good reduction over finite fields.

In this article, we investigate the discrete Painlev\'{e} II equation (dP\II) over finite fields.
The discrete Painlev\'{e} equations are non-autonomous, integrable mappings which tend to some continuous Painlev\'{e} equations for appropriate choices of the continuous limit \cite{RGH}.
They do not have good reduction modulo prime in general.
However we will show that they have \textit{almost good reduction} the precise meaning of which will be given later.
We show that the time evolution of the discrete  Painlev\'{e} equations can be well defined generically, using the reduction from a local field $\Q_p$ to a finite field $\F_p$. This reduction is shown to be well-defined and is used to obtain some special solutions directly from those over fields of characteristic zero such as $\Q$ or $\R$.
%

%
%
\section{The dP\II equation over a local field and its reduction modulo prime}
%
A discrete Painlev\'{e} equation is a non-autonomous and nonlinear second order ordinary difference equation with several parameters.
When it is defined over a finite field, the dependent variable takes only a finite number of values and its time evolution will attain an indeterminate state for generic values of the parameters and initial conditions.
The dP\II equation is defined as 
\begin{equation}
u_{n+1}+u_{n-1}=\frac{z_n u_n+a}{1-u_n^2}\quad (n \in \mathbb{Z}),
\label{dP2equation}
\end{equation}
where $z_n=\delta n + z_0$ and $a, \delta, z_0$ are constant parameters \cite{NP}.
When (\ref{dP2equation}) is defined over a finite field $\F_p$,
the dependent variable $u_n$ will eventually take values $\pm 1$ for generic parameters and initial values $(u_0,u_1) \in \F_{p}^2$, 
and we cannot proceed to evolve it.
To determine its time evolution consistently, we have two choices:
One is to restrict the parameters and the initial values to a smaller domain so that the singularities do not appear.
The other is to extend the domain on which the equation is defined, which will be adopted in this paper.
It is convenient to rewrite (\ref{dP2equation}) as:
\begin{equation}
\left\{
\begin{array}{cl}
x_{n+1}&=\displaystyle\frac{\alpha_n}{1-x_n}+\displaystyle\frac{\beta_n}{1+x_n}-y_{n},\\
y_{n+1}&=x_n,
\end{array}
\right.
\label{dP2}
\end{equation}
where $\alpha_n:=\frac{1}{2}(z_n+a),\ \beta_n:=\frac{1}{2}(-z_n+a)$.
Let $p$ be a prime number and for each $x \in \Q$ ($x \neq 0$) write $x=p^{v_p(x)} \displaystyle\frac{u}{v}$ where $v_p(x), u, v \in \Z$ and $u$ and $v$ are coprime integers neither of which is divisible by $p$.
The $p$-adic norm $|x|_p$ is defined as $|x|_p=p^{-v_p(x)}$. ($|0|_p=0$.)
The local field $\Q_p$ is a completion of $\Q$ with respect to the $p$-adic norm. 
It is called the field of $p$-adic numbers and its subring $\Z_p:=\{x\in \Q_p | \ |x|_p \le 1\}$ is called the ring of $p$-adic integers. 
The $p$-adic norm satisfies a sharper triangle inequality 
\begin{equation}
|x+y|_p \le \max[|x|_p,|y|_p ],
\label{ptri}
\end{equation} 
where equality holds whenever $|x|_p \ne |y|_p$.
Let $\mathfrak{p}$ be the maximal ideal of $\Z_p$,
\[
\mathfrak{p}:=\left\{x \in \Z_p |\ v_p(x) \ge 1 \right\}. 
\]
We define $\tilde{x}$ as the reduction of $x$ modulo $\mathfrak{p}$: $x \in \Z_p \to \tilde{x} \in \Z_p/\mathfrak{p} \cong \F_p$. We easily show that this reduction is a surjective ring homomorphism:
\begin{equation}
\widetilde{x \pm y}=\tilde{x} \pm \tilde{y},\quad \widetilde{x \cdot y}=\tilde{x} \cdot \tilde{y}, \label{prel}
\end{equation}
for $x,y \in \Z_p$.
For a rational map $\phi$: $\mathcal{D} \subseteq \Z_p^2 \to \Z_p^2$, when it is expressed  on some domain $\mathcal{D}$ as
\[
\phi(x,y)=\frac{\sum_{i_1,i_2 \ge 0}a_{i_1i_2}x^{i_1}y^{i_2}}{\sum_{j_1,j_2 \ge 0}b_{j_1j_2}x^{j_1}y^{j_2}}\ \in\mathbb{Z}_p(x,y),
\]
coefficient reduction $\tilde{\phi}$ is defined by
\[
\tilde{\phi}(x,y)=\frac{\sum_{i_1,i_2 \ge 0}\widetilde{a_{i_1i_2}}x^{i_1}y^{i_2}}{\sum_{j_1,j_2 \ge 0}\widetilde{b_{j_1j_2}}x^{j_1}y^{j_2}}\ \in\mathbb{F}_p(x,y).
\]
The rational map $\phi$ is said to have \textit{good reduction} (modulo $\mathfrak{p}$ on the domain $\mathcal{D}$) if it holds that $\widetilde{\phi(x,y)}=\tilde{\phi}(\tilde{x},\tilde{y})$ for any $(x,y) \in \mathcal{D}$ \cite{Silverman}.
We define a generalized notion;
\begin{Definition}
A (non-autonomous) rational map $\phi_n$: $\mathcal{D} \subseteq \Z_p^2 \to \Q_p$ $(n \in \Z)$ is said to have almost good reduction modulo $\mathfrak{p}$ if there
exists a positive integer $m_{\mbox{\rm \scriptsize p};n}$ for any $\mbox{\rm p}=(x,y) \in \mathcal{D}$ and time step $n$ such that
\begin{equation}
\widetilde{\phi_n^{m_{\mbox{\rm \tiny p};n}}(x,y)}=\widetilde{\phi_n^{m_{\mbox{\rm \tiny p};n}}}(\tilde{x},\tilde{y}),
\label{AGR}
\end{equation}
where $\phi_n^m :=\phi_{n+m-1} \circ \phi_{n+m-2} \circ \cdots \circ \phi_n$.
\end{Definition} 

To see the significance of this notion of \textit{almost good reduction}, let us consider the mapping $\Psi_\gamma$:
\begin{equation}
\left\{
\begin{array}{cl}
x_{n+1}&=\displaystyle\frac{ax_n+1}{x_n^\gamma y_n}\\
y_{n+1}&=x_n
\end{array}
\right.,
\label{discretemap}
\end{equation} 
where $a \in \{1,2,\cdots, p-1\}$  and $\gamma \in \Z_{\ge 0}$ are parameters. 
The map (\ref{discretemap}) is known to be integrable if and only if $\gamma=0,1,2$.
When $\gamma=1, 2$, (\ref{discretemap}) belongs to the QRT family and is integrable in the sense that it has a conserved quantity.
Let $\mathcal{D}$ be the domain $\{(x,y) \in \Z_p \ |x \ne 0, y \ne 0\}$, then clearly 
\[
\widetilde{\Psi_2(x_n,y_n)}=\widetilde{\Psi}_2(\tilde{x}_n,\tilde{y}_n) \qquad \mbox{for $\tilde{x}_n \ne 0, \ \tilde{y}_n \ne 0$}.
\] 
For $(x_n,y_n)\in\mathcal{D}$ with $\tilde{x}_n=0$ and $\tilde{y}_n \ne 0$, we find that
$\widetilde{\Psi_2^k}(\tilde{x}_n=0,\tilde{y}_n)$ is not defined for $k=1,2$,
however it is defined if $k=3$ and we have
\[
\widetilde{\Psi_2^3(x_n,y_n)}=\widetilde{\Psi_2^3}(\tilde{x}_n=0,\tilde{y}_n)=\left(\displaystyle\frac{1}{a^2\tilde{y}},0\right) .
\]
Finally for $\tilde{x}_n=\tilde{y}_n = 0$, we find that
$\widetilde{\Psi_2^k}(\tilde{x}_n,\tilde{y}_n)$ is not defined for $k=1,2,..,7$,
however
\[
\widetilde{\Psi_2^8(x_n,y_n)}=\widetilde{\Psi_2^8}(\tilde{x}_n=0,\tilde{y}_n=0)=\left(0,0\right) .
\]
Hence the map $\Psi_2$ has almost good reduction modulo $\mathfrak{p}$ on $\mathcal{D}$.
Note that, in the case $\gamma=2$ and $a=0$,
if we take 
\[
f_{2k}:=x_{2k}x_{2k-1},\ f_{2k-1}:=\frac{1}{x_{2k-1}x_{2k-2}}
\]
(\ref{discretemap}) turns into the trivial linear mapping $f_{n+1}=f_n$ which has apparently good reduction modulo $\mathfrak{p}$.
In a similar manner, we find that $\Psi_\gamma$ ($\gamma=0,1$) also has almost good reduction modulo $\mathfrak{p}$ on $\mathcal{D}$. 
On the other hand, for $\gamma \ge 3$ and $\tilde{x}_n=0$, we easily find that
\[
{}^\forall k \in \Z_{\ge 0}, \;\; \widetilde{\Psi_\gamma^{k}(x_n,y_n)} \ne \widetilde{\Psi_\gamma^{k}}(\tilde{x}_n=0,\tilde{y}_n),
\]
since the order of $p$ diverges as we iterate the mapping.
Thus we have proved the following proposition:
\begin{Proposition}
The rational mapping (\ref{discretemap}) has almost good reduction modulo $\mathfrak{p}$ only for $\gamma=0,1,2$.
\label{PropQRT}
\end{Proposition}
Note that having almost good reduction is equivalent to the integrability of the equation in these examples.
 
Now let us examine the dP\II equation (\ref{dP2}) over $\Q_p$. 
We suppose that $p \ge 3$, and redefine the coefficients $\alpha_n$ and $\beta_n$ so that
they are periodic with period $p$:
\begin{eqnarray*}
\alpha_{i+mp}:=\frac{(i\delta+z_0+a+n_\alpha p)}{2},\ \beta_{i+mp}:=\frac{(-i\delta-z_0+a+n_\beta p)}
{2},\\
(m\in\Z,\ i\in \{0,1,2,\cdots,p-1\}),
\end{eqnarray*}
where the integer $n_\alpha$ ($n_\beta$) is chosen such that $0 \in \{\alpha_i\}_{i=0}^{p-1}$ $(0 \in \{\beta_i\}_{i=0}^{p-1})$.
As a result, we have $\tilde{\alpha}_{n}=\widetilde{\frac{n \delta +z_0+a}{2}}$, $\tilde{\beta}_{n}=\widetilde{\frac{-n \delta -z_0+a}{2}}$ and $|\alpha_n|_p,\ |\beta_n|_p\in\{0,1\}$ for any integer $n$.

\begin{Proposition}
Under the above assumptions, the dP\II equation has almost good reduction modulo $\mathfrak{p}$ on $\mathcal{D}:=\{ (x,y) \in \Z_p^2\ |x \ne \pm 1\}$.
\label{PropdP2}
\end{Proposition}
\textbf{Proof}

We put $(x_{n+1},y_{n+1})=\phi_n(x_n,y_n)=\left( \phi_n^{(x)}(x_n,y_n),\phi_n^{(y)} (x_n,y_n) \right)$.
When $\tilde{x}_n \ne \pm 1$, we have from (\ref{prel})
\[
\tilde{x}_{n+1}=\displaystyle\frac{\tilde{\alpha_n}}{1-\tilde{x}_n}+\displaystyle\frac{\tilde{\beta_n}}{1+\tilde{x}_n}-\tilde{y}_{n},
\quad \tilde{y}_{n+1}=\tilde{x}_n.
\]
Hence $\widetilde{\phi_n(x_n,y_n)}=\tilde{\phi}_n(\tilde{x}_n,\tilde{y}_n)$.

When $\tilde{x}_n=1$, we can write $x_n=1+p^k e$ $(k \in \Z_+,\ |e|_p=1)$. 
We have to consider four cases:\\
\noindent
(i) For $\alpha_n= 0 $,
\[
\tilde{x}_{n+1}=\tilde{\phi}_n^{(x)}(\tilde{x}_n,\tilde{y}_n)=\widetilde{\left(\frac{\beta_n}{2}\right)}-\tilde{y}_n.
\]
Hence we have $\widetilde{\phi_n(x_n,y_n)}=\tilde{\phi}_n(\tilde{x}_n,\tilde{y}_n)$.\\
(ii) In the case $\alpha_n \neq 0$ and $\beta_{n+2} \neq 0$, 
\begin{eqnarray*}
x_{n+1}&=&-\displaystyle\frac{(\alpha_n-\beta_n)(1+ep^k)+a}{ep^k(2+ep^k)}-y_n=-\displaystyle\frac{2\alpha_n+(\alpha_n-\beta_n)ep^k}{ep^k(2+ep^k)}-y_n,\\
x_{n+2}&=&-\frac{\alpha_n^2+\mbox{polynomial of $O(p)$}}{\alpha_n^2+\mbox{polynomial of $O(p)$}},\\
x_{n+3}&=&\displaystyle\frac{\{2\alpha_{n}y_n+2\delta \beta_{n+1}+(2-\delta)a \}\alpha_n^3 +\mbox{polynomial of $O(p)$}}{2\beta_{n+2}
\alpha_n^3 + \mbox{polynomial of $O(p)$} }.
\end{eqnarray*}
Thus we have
\[
\tilde{x}_{n+3}=\frac{2\tilde{\alpha}_{n}\tilde{y}_n+2\delta \tilde{\beta}_{n+1}+(2-\delta)a}{2 \tilde{\beta}_{n+2}},
\quad \tilde{y}_{n+3}=-1,
\]
and $\widetilde{\phi_n^3(x_n,y_n)}=\widetilde{\phi_n^3}(\tilde{x}_n,\tilde{y}_n)$.\\
(iii) In the case $\alpha_n \neq 0$, $\beta_{n+2}= 0$ and $a \ne -\delta$, we have to calculate up to $x_{n+5}$.
After a lengthy calculation we find
\begin{eqnarray*}
\tilde{x}_{n+4}&=&\widetilde{\phi_n^{5}}^{(y)}(1,\tilde{y}_n)=1,\\
\tilde{x}_{n+5}&=&\widetilde{\phi_n^{5}}^{(x)}(1,\tilde{y}_n)=-\frac{a\delta-(a-\delta)\tilde{y}_n}{a+\delta},
\end{eqnarray*}  
and we obtain $\widetilde{\phi_n^5(x_n,y_n)}=\widetilde{\phi_n^5}(\tilde{x}_n,\tilde{y}_n)$.\\
(iv) Finally, in the case $\alpha_n \neq 0$, $\beta_{n+2}= 0$ and $a = -\delta$ we have to calculate up to $x_{n+7}$.
The result is
\begin{eqnarray*}
\tilde{x}_{n+6}&=&\widetilde{\phi_n^{7}}^{(y)}(1,\tilde{y}_n)=-1,\\
\tilde{x}_{n+7}&=&\widetilde{\phi_n^{7}}^{(x)}(1,\tilde{y}_n)=\frac{1+2\tilde{y}_n}{2},
\end{eqnarray*}
and we obtain $\widetilde{\phi_n^7(x_n,y_n)}=\widetilde{\phi_n^7}(\tilde{x}_n,\tilde{y}_n)$.
Hence we have proven that the dP\II equation has almost good reduction modulo $\mathfrak{p}$ at $\tilde{x}_n=1$.

We can proceed in the case $\tilde{x}_n=-1$ in an exactly similar manner.\qed

From this proposition, the evolution of the dP\II equation (\ref{dP2equation}) over $\P\F_p$ can be constructed from the following seven cases which determine $\{u_{n+1},u_{n+2},\cdots\}$ from the initial values $u_{n-1}$ and $u_n$.
Note that we can assume that $u_{n-1} \ne \infty$ because all the cases in which the dependent variable $u_n$ becomes $\infty$ are included below.
\begin{enumerate}
\item For $u_n \in \{2,3,...,p-2\}$, or $u_n=1$ and $\alpha_n = 0$, or $u_n=p-1$ and $\beta_n =0$,
\[
u_{n+1}=\frac{{\alpha}_n}{1-u_n}+\frac{{\beta}_n}{1+u_n}-u_{n-1}.\\
\]
\item For $u_n=1$, $\alpha_n \ne 0$ and $\beta_{n+2} \ne 0$,
\[
u_{n+1}=\infty,\  u_{n+2}=p-1,\  u_{n+3}=\frac{2\alpha_n u_{n-1}+2\delta\beta_{n+1}+(2-\delta)a}{2\beta_{n+2}}.
\]
\item For $u_n=1$, $\alpha_n \ne 0$, $\beta_{n+2} = 0$ and $a+\delta \ne 0$,
\begin{eqnarray*}
&u_{n+1}=\infty,\  u_{n+2}=p-1,\  u_{n+3}=\infty,\\
&\qquad u_{n+4}=1,\ u_{n+5}=\displaystyle-\frac{a\delta-(a-\delta){u}_{n-1}}{a+\delta}.
\end{eqnarray*}
\item For $u_n=1$, $\alpha_n \ne 0$, $\beta_{n+2} = 0$ and $a+\delta = 0$,
\begin{eqnarray*}
&u_{n+1}=\infty,\  u_{n+2}=p-1,\  u_{n+3}=\infty,\ u_{n+4}=1,\ u_{n+5}=\infty,\\
&\qquad u_{n+6}=p-1,\ u_{n+7}=\displaystyle\frac{1+2{u}_{n-1}}{2}.
\end{eqnarray*}
\item For $u_n=p-1$, $\beta_n \ne 0$ and $\alpha_{n+2} \ne 0$,
\[
u_{n+1}=\infty,\ u_{n+2}=1,\ u_{n+3}=\frac{a(-2+\delta)-2\delta\alpha_{n+1}+2\beta_n {u}_{n-1}}{2 \alpha_{n+2}}.
\]
\item For $u_n=p-1$, $\beta_n \ne 0$, $\alpha_{n+2} = 0$ and $a \ne \delta$,
\begin{eqnarray*}
&u_{n+1}=\infty,\ u_{n+2}=1,\ u_{n+3}=\infty,\\
&\qquad u_{n+4}=p-1,\ u_{n+5}=\displaystyle\frac{a \delta+(a+\delta){u}_{n-1}}{a-\delta}.
\end{eqnarray*}
\item For $u_n=p-1$, $\beta_n \ne 0$, $\alpha_{n+2} = 0$ and $a = \delta$,
\begin{eqnarray*} 
&u_{n+1}=\infty,\ u_{n+2}=1,\ u_{n+3}=\infty,\ u_{n+4}=p-1,\ u_{n+5}=\infty, \\
&\qquad u_{n+6}=1,\ u_{n+7}=\displaystyle\frac{-1+2{u}_{n-1}}{2}.
\end{eqnarray*}
\end{enumerate}
The above approach is closely related to the singularity confinement method which is an effective test to judge the integrability of the given equations \cite{Grammaticosetal}.
In the proof of the Proposition \ref{PropdP2}, we take $x_n=1+e p^k$ and show that the limit $\lim_{|e p^k|_p \to 0}(x_{n+m}, x_{n+m+1})$ 
is well defined for some positive integer $m$.
Here $ep^k$ is an alternative in $\Q_p$ for the infinitesimal parameter $\epsilon$ in the singularity confinement test in $\C$.
From this observation and propositions \ref{PropQRT} and \ref{PropdP2}, we postulate that having almost good reduction in arithmetic mappings is similar to passing the singularity confinement test.

Now we consider special solutions to (\ref{dP2equation}) over $\P\F_p$.
For the dP\II equation over $\C$, rational function solutions have already been obtained \cite{Kajiwara}.
Let $N$ be a positive integer and $\lambda \ne 0$  be a constant. Suppose that 
\[
L_k^{(\nu)}(\lambda):=\left\{ 
\begin{array}{cl}
\displaystyle
\sum_{r=0}^k(-1)^r
\left(
\begin{array}{cl}
k+\nu\\
k-r
\end{array}
\right) 
\displaystyle\frac{\lambda^r}{r!}&\quad(k \in \Z_{\ge 0}),\\
0 &\quad (k \in \Z_{<0}),
\end{array}
\right.
\]
and
\begin{equation}
\tau_N^n:=
\left|
\begin{array}{cccc}
L_N^{(n)}(\lambda)&L_{N+1}^{(n)}(\lambda)&\cdots&L_{2N-1}^{(n)}(\lambda)\\
L_{N-2}^{(n)}(\lambda)&L_{N-1}^{(n)}(\lambda)&\cdots&L_{2N-3}^{(n)}(\lambda)\\
\vdots&\vdots &\ddots &\vdots\\
L_{-N+2}^{(n)}(\lambda)&L_{-N+3}^{(n)}(\lambda)&\cdots&L_{1}^{(n)}(\lambda)
\end{array}\right|,
\label{Ltau}
\end{equation}
where $a=-\frac{2(N+1)}{\lambda}$ and $\delta=z_0=\frac{2}{\lambda}$.
Then a rational function solution of the dP\II equation is given by
\begin{equation}
u_n=\frac{\tau_{N+1}^{n+1}\tau_{N}^{n-1}}{\tau_{N+1}^n\tau_N^n}-1.
\label{rationaldP2}
\end{equation}
If we deal with the terms in (\ref{Ltau}) and (\ref{rationaldP2}) by arithmetic operations over $\F_p$, 
we encounter terms such as $\frac{p^k}{p^l}$ and (\ref{rationaldP2}) is not well-defined.
However, from proposition \ref{PropdP2}, we find that (\ref{rationaldP2}) gives a solution to the dP\II equation over $\P\F_q$ by
the reduction from $\Q (\subset \Q_p)$, as long as the solution avoids the points $(\tilde{\alpha}_n=0,\ u_n=1)$ and $(\tilde{\beta}_n=0,\ u_n=-1)$, which is equivalent to the solution satisfying
\begin{equation}
\tau_{N+1}^{-N-1} \tau_N^{-N-3}\not\equiv 0,\ \frac{\tau_{N+1}^{N+1}\tau_N^{N-1}}{\tau_{N+1}^N\tau_N^N}\not\equiv 2, \label{taucond}
\end{equation}
where the superscripts are considered modulo $p$. Note that $\tau_N^n\equiv \tau_N^{n+p}$ for all integers $N$ and $n$.
In the table below, we give several \textit{rational solutions to the dP\II equation} with $N=3$ and $\lambda=1$ over $\P\F_q$ for $q=3,5,7$ and $11$. We see that the period of the solution is $p$.
\[
\begin{array}{|c|c|c|l|}
\hline
& & & \\[-2mm]
\raise5mm\hbox{$p$} & \raise5mm\hbox{\small{$\tau_{N+1}^{-N-1}\tau_N^{-N-3}$}} & \raise5mm\hbox{$\frac{\tau_{N+1}^{N+1}\tau_N^{N-1}}{\tau_{N+1}^N\tau_N^N}$}
&\enskip
\raise5mm\hbox{$u_1,u_2,u_3,u_4,u_5,u_6,u_7,u_8,u_9,u_{10},\ldots$} \\ \hline & & & \\[-3mm]
3 & \infty & \infty & \
\raise3mm\hbox{$\underbrace{1,2,\infty}_{\mbox{period
$3$}},1,2,\infty,1,2,\infty,1,2,\infty,1,2,\ldots$}
\\[4mm] \hline
& & & \\[-3mm]
5 & \infty & 4 &\
\raise3mm\hbox{$\underbrace{4,2,3,1,\infty}_{\mbox{period
$5$}},4,2,3,1,\infty,4,2,3,1,\infty,\ldots$} \\[4mm] \hline & & & \\[-3mm]
7 & \infty & 0 &\
\raise3mm\hbox{$\underbrace{1,\infty,6,5,1,\infty,6}_{\mbox{period
$7$}},1,\infty,6,5,1,\infty,6,1,\ldots$} \\[4mm] \hline & & & \\[-3mm]
11 & 0 & 7 &\
\raise3mm\hbox{$\underbrace{\infty,1,6,1,\infty,10,\infty,1,0,2,10}_{\mbox{period
$11$}},\infty,1,6,\ldots$} \\[4mm] \hline
\end{array} \]
We see from the case of $p=11$ that we may have an appropriate solution even if the condition (\ref{taucond}) is not satisfied, although this is not always true.
The dP\II equation has linearized solutions also for $\delta=2a$ \cite{Tamizhmani}. 
With our new method, we can obtain the corresponding solutions without difficulty.
Our method of almost good reduction is expected to serve as a criterion for integrability of the discrete systems over finite fields.

\section{Concluding remarks}
In this article we investigated the discrete Painlev\'{e} II equation over finite fields.
To avoid indeterminacy, we examined the reduction modulo prime number from a $p$-adic number field $\Q_p$.
We defined the notion of \textit{almost good reduction} which is an arithmetic analogue of passing the singularity confinement test, and proved that the discrete Painlev\'{e} II equation has this property.
Thanks to this property, not only the time evolution of the discrete Painlev\'{e} equations can be well defined, but also
a solution over $\Q$ or $\Q_p$ can be directly transferred to a solution over $\P\F_p$.
We presented the special solutions over $\P\F_p$.
Although we examined only an example of the QRT family and the dP\II equation in this paper, we conjecture that this approach is equally valid in other discrete Painlev\'{e} equations and its generalisation \cite{KNY}. 
Furthermore, we expect that this `almost good reduction' criterion can be applied to finding higher order \textit{integrable} mappings in arithmetic dynamics, and
that a similar approach is also useful for the investigation of discrete partial difference equations such as soliton equations over finite fields \cite{DBK,KMT}. These problems are currently being investigated.      

\section*{Acknowledgement}
The authors wish to thank Prof. R. Willox for useful comments.
This work is partially supported by Grant-in-Aid for Scientific Research of Japan Society for the Promotion of Science ($24\cdot 1379$).

\end{document}